# Sorting polarized photons with a single scatterer

Francisco J. Rodríguez-Fortuño, Daniel Puerto, Amadeu Griol, Laurent Bellieres, Javier Martí and Alejandro Martínez*

Nanophotonics Technology Center, Universitat Politècnica de València, Valencia (Spain)

*Correspondence to:  amartinez@ntc.upv.es

**Abstract:** Intuitively, light impinging on a spatially symmetric object will be scattered symmetrically. This intuition can fail at the nanoscale if the polarization of the incoming light is properly tailored. In fact, it has been demonstrated that near-field interference can result in the unidirectional excitation of plasmonic modes using circularly polarized light. Here we show that linearly-polarized photons impinging on a single spatially-symmetric scatterer created in a silicon waveguide are guided into a certain direction of the waveguide depending exclusively on their polarization angle. Reciprocity implies that the scatterer can also act as a two-input dielectric nanoantenna radiating two different linear polarizations. Our work broadens the scope of near-field interference beyond plasmonics, with applications in polarization (de)multiplexing, unidirectional coupling, directional switching, radiation polarization control, and polarization-encoded quantum information processing in photonic integrated circuits.

**Main text:**

Wave interference is a key phenomenon behind many applications in optics. In most cases, interference occurs between propagating waves accumulating different phase shifts, as in the case of interferometers or phased-array antennas [1,2]. However, wave interference can also take place in the near field, where the distances are so small that phase shifts can be neglected. For instance, near-field radiationless interference has been used to achieve subwavelength resolution in imaging [3] as well as in nanoplasmonic systems as a way to switch between propagation paths or localized spots [4–8]. More recently, a series of works have demonstrated that illuminating a plasmonic structure with circularly polarized light can result in unidirectional guidance of surface plasmons [9–13]. This novel and somewhat unexpected property is given different interpretations, but ultimately relies in near field interference and superposition in the excitation of waves. This powerful concept has been demonstrated for plasmonic waves, but it is so fundamental that it can be extended to any class of waveguide [9].

In this work we use near-field interference to sort linearly-polarized photons impinging normally on a mirror-symmetric scatterer built into a standard silicon waveguide so that incoming photons can be directed towards each waveguide direction depending exclusively on their polarization. In comparison with previous experiments showing directional excitation of surface plasmons [9–13], this design works with linear polarization so a quarter-wave plate is not needed, and the photons are directly coupled to dielectric waveguide structures, thus greatly simplifying the experimental scheme and potential practical realizations. The technique is non-resonant although there is a relationship between the wavelength and the polarization angle required to direct photons towards opposite directions. Our experiments were performed in the 1,550 nm wavelength regime and using silicon waveguides but the phenomenon could be easily extrapolated to other wavelengths and materials. This scheme is by far simpler than other configurations of near field interference that require complex time-varying phase and amplitude polarization patterns to control the contrast between the optical field confined in nanoparticles [7] or guided towards different spatial regions [4]. In our case we only have to vary the angle of the incident linear polarization of monochromatic light. All previous experiments showing control of light localization and directional guiding using near-field interference have been done in plasmonics, owing to its ability for subwavelength confinement. Here, we present evidence of this type of phenomenon in conventional dielectric integrated photonics exploiting the near field excitation of guided modes, opening the field to a broader range of scientists and technologists in nanooptics. It deserves to be mentioned that two dimensional grating couplers have been used in silicon photonics as efficient polarization splitters [14–17], but their underlying physical phenomenon is completely different: the output light for different incoming polarizations exit at 90º to each other, and opposite outputs are excited by the same polarization due to symmetry. Tilted illumination can concentrate the power into one direction [17], but here we achieve unidirectional propagation using normal incidence based on local near field interference.

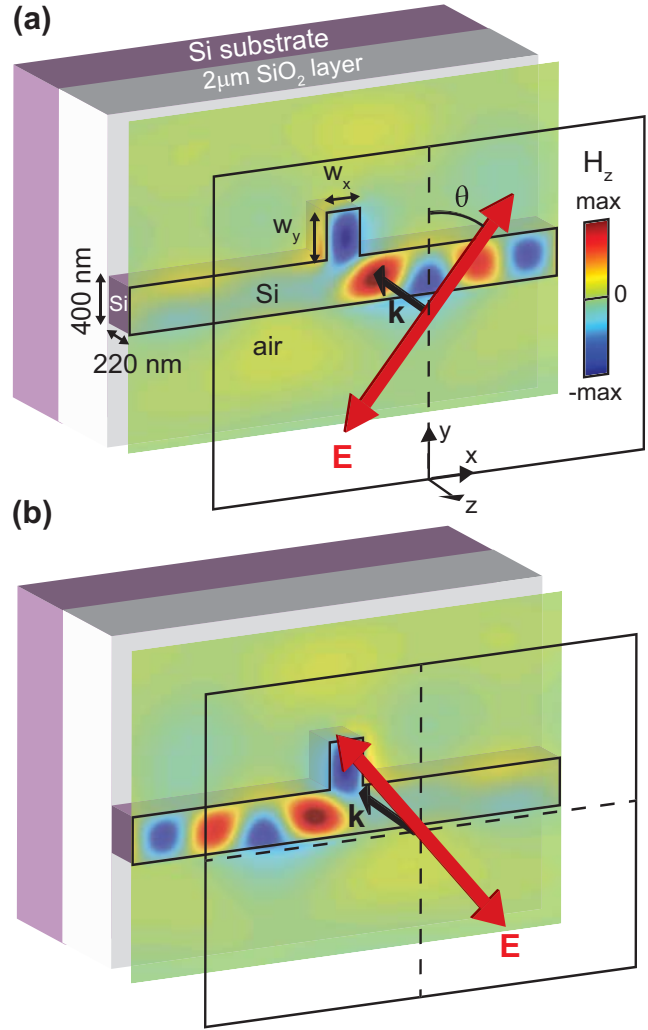

**Fig. 1**. **Sorting of polarized photons in a silicon waveguide with a single scatterer:** Numerical simulation showing the instantaneous magnetic field z-component at the middle plane of the waveguide for a normally incident, linearly polarized monochromatic $\lambda = 1583\,\text{nm}$ plane wave with a polarization (**a**) $\theta = +40°$ and (**b**) $\theta = -40°$ achieving an ideal 1:0 contrast ratio.

Figure 1 shows the basic idea. Linearly polarized light impinges on a scatterer (we designed a rectangular structure with sides $w_x = 300\,\text{nm}$ and $w_y = 400\,\text{nm}$) fabricated on a silicon waveguide of standard dimensions ($220 \times 400\,\text{nm}$). The structure scatters light, which in the near field couples to the fundamental transverse electric (TE) mode of the silicon waveguide. For a linear polarization at an angle $\theta$ with the y axis, the near field vectorial interference [9] of such scattering results in the excitation of the waveguide modes only on one direction. A possible explanation is given by the relative phases in which the longitudinal and transverse components of the electric field of the waveguide mode are excited by the scattering [9]. The phenomenon can also be understood by studying the wavevector components of the scattered field [9], as well as through magnetic induction currents [12]. Most practically in this case, an explanation can be given using the superposition principle, by studying the near field interference in the scattering of horizontally- and vertically-polarized components of the incident light [10], as described below.

General symmetry considerations imply that, for any possible horizontally mirror-symmetric scatterer, an incident vertical polarization ($\mathbf{E}_{inc} = E_0 e^{ikz} \hat{\mathbf{y}}$) excites the guided mode with complex amplitudes +A to the right (+x) and −A to the left (−x) [see Fig 2(a)], while a horizontal polarization ($\mathbf{E}_{inc} = E_0 e^{ikz} \hat{\mathbf{x}}$) excites both directions of the guided modes with the same complex amplitude B [see Fig. 2(b)]. Applying the superposition principle, the interference between fields with an even and odd x-symmetry results in the unidirectionality shown in Fig. 1.

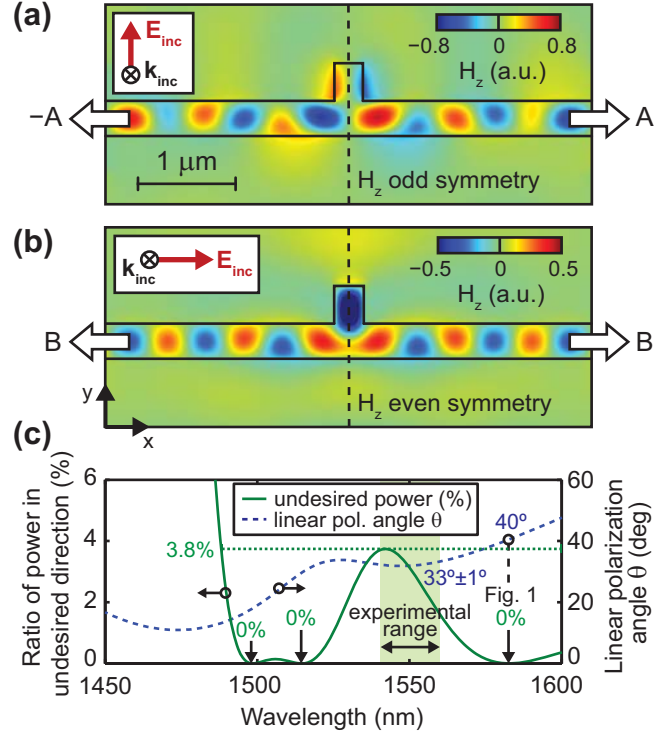

**Fig. 2. Symmetry explanation of the phenomenon and scatterer performance:** (a-b) Numerical simulation showing the instantaneous $H_z$ at the middle plane of the waveguide for a normally incident, linearly polarized monochromatic $\lambda = 1583$ nm plane wave for (a) vertical polarization $\mathbf{E}_{inc} = E_0 e^{ikz} \hat{\mathbf{y}}$ ($\theta = 0°$) and (b) horizontal polarization $\mathbf{E}_{inc} = E_0 e^{ikz} \hat{\mathbf{x}}$ ($\theta = 90°$). (C) Ratio of power between the undesired and the desired direction $R = (A - (\beta/\alpha)B)/(A + (\beta/\alpha)B)$ as a function of wavelength. In general, R could be made zero everywhere if the appropriate elliptical polarization $\beta/\alpha = A/B$ was considered, but here we limit ourselves to linear polarization only $(\beta/\alpha) \in \Re$ so the optimum (minimum) contrast R is obtained when $(\beta/\alpha) = |A/B| \cos(\arg(A/B))$ (easily demonstrated from geometric arguments in the complex plane). The associated linear polarization angle given by $\theta = \tan^{-1}(\beta/\alpha)$ is also plotted.

Mathematically, any normally incident plane wave can be written as $\mathbf{E}_{inc} = E_0 e^{ikz} (\alpha \hat{\mathbf{y}} + \beta \hat{\mathbf{x}})$, where α and β are complex coefficients determining the polarization. Applying the superposition principle, this general situation corresponds to a simple weighted addition of the fields in Figs. 2(a) and 2(b), so that the waveguide TE mode will be excited with amplitudes $\beta B + \alpha A$ to the right side and $\beta B - \alpha A$ to the left. A perfect unidirectionality with ratio 1:0 can be achieved under the simple condition $\beta = (A/B)\alpha$, where all quantities are complex, and it requires in general an elliptical polarization. Changing $\beta \to -\beta$ switches to the opposite propagation direction [5], as shown in Fig. 1. For all this to hold, the scatterer must not be symmetric in the y-direction, otherwise $B = 0$ (i.e. the incoming H polarization does not excite the TE waveguide mode for a y-symmetric scatterer). Thus, breaking the symmetry in the y-axis is a necessary fundamental requirement to achieve asymmetrical excitation in the x-axis. We use a y-symmetric structure in our experiments (see Fig. 3) as a control scatterer which shows a completely symmetric power splitting with no unidirectionality.

In our design, we want to achieve unidirectionality when the incident polarization is linear, i.e. $\alpha$ and $\beta$ have equal phase. Under this condition, ideal unidirectionality requires that the phases of A and B are equal. A and B can be easily obtained from numerical simulations and show a dispersive behaviour. At the wavelengths where their phase is equal, a linear polarization will achieve theoretically a perfect 1:0 contrast. In our design, A and B have a very similar phase (< 22° difference) throughout the entire range of telecommunication wavelengths, meaning that the ideal 1:0 contrast condition would require a very eccentric, close to linear, incoming polarization ellipse. In practice, as shown by the simulation and experiments, an incoming linear polarization is sufficient for high contrast ratios in that range. Fig. 2(c) plots the ratio of power R between the undesired and the desired direction when using the optimal linear polarization. The points where R = 0% correspond to frequencies in which the phases of A and B are equal. The angle $\theta$ of the linear polarization required to maximize the contrast is also plotted in Fig 2(c), and depends directly on A and B, so optimization algorithms are easily applied to the scatterer design. In our case, at a wavelength of 1583 nm we theoretically achieve ideal 1:0 unidirectionality at ±40° (plotted in Fig. 1). In the wide region around telecommunication C-band wavelengths which we measured, 1540 nm – 1560 nm, we theoretically predict a unidirectionality ratio always better than 25:1 and a very stable polarization angle $\theta \approx 33°$ of photon sorting, with variations smaller than 1° in the entire range.

Our designed scatterer can also be regarded as a receiving antenna with an effective area (defined as the ratio between the power excited in the waveguide and the incident power density) of ~3200 nm$^2$ (at 1560 nm), which is 27% of its physical area. This allows a very simple experimental realization. Efficient improved scatterers using metals can be designed [18].

A depiction of our experimental setup is shown in Fig 3(a). A lensed fibre is used to illuminate five different scatterers with linearly polarized light. One of them is our control scatterer, symmetric in the y direction. The others are two pairs of scatterers with the same dimensions but mirrored in y, thus exhibiting opposite directionality. Fig 3(b) shows a scanning electron micrograph of the fabricated structure. The light excited into each waveguide is guided toward the edge of the sample, where the waveguides are cut and radiate light, through a pinhole (to avoid collecting noise light from the nearby lensed fibre) into a microscope objective with an IR camera which can image the intensity of the 5 different outputs. Figs. 3(c-e) shows the obtained IR images at 1550 nm for three different fibre illumination polarizations, demonstrating clearly the polarization sorting of photons. For further quantitative proof, Fig 3(f) shows the different spot powers at 1550 nm for different input polarization angles in steps of 5 degrees, compared with the expected results obtained from simulations. The graph for wavelengths 1540 nm and 1560 nm are given in the Supplemental Material [19], Fig. S1, showing the same effect in a broad bandwidth.

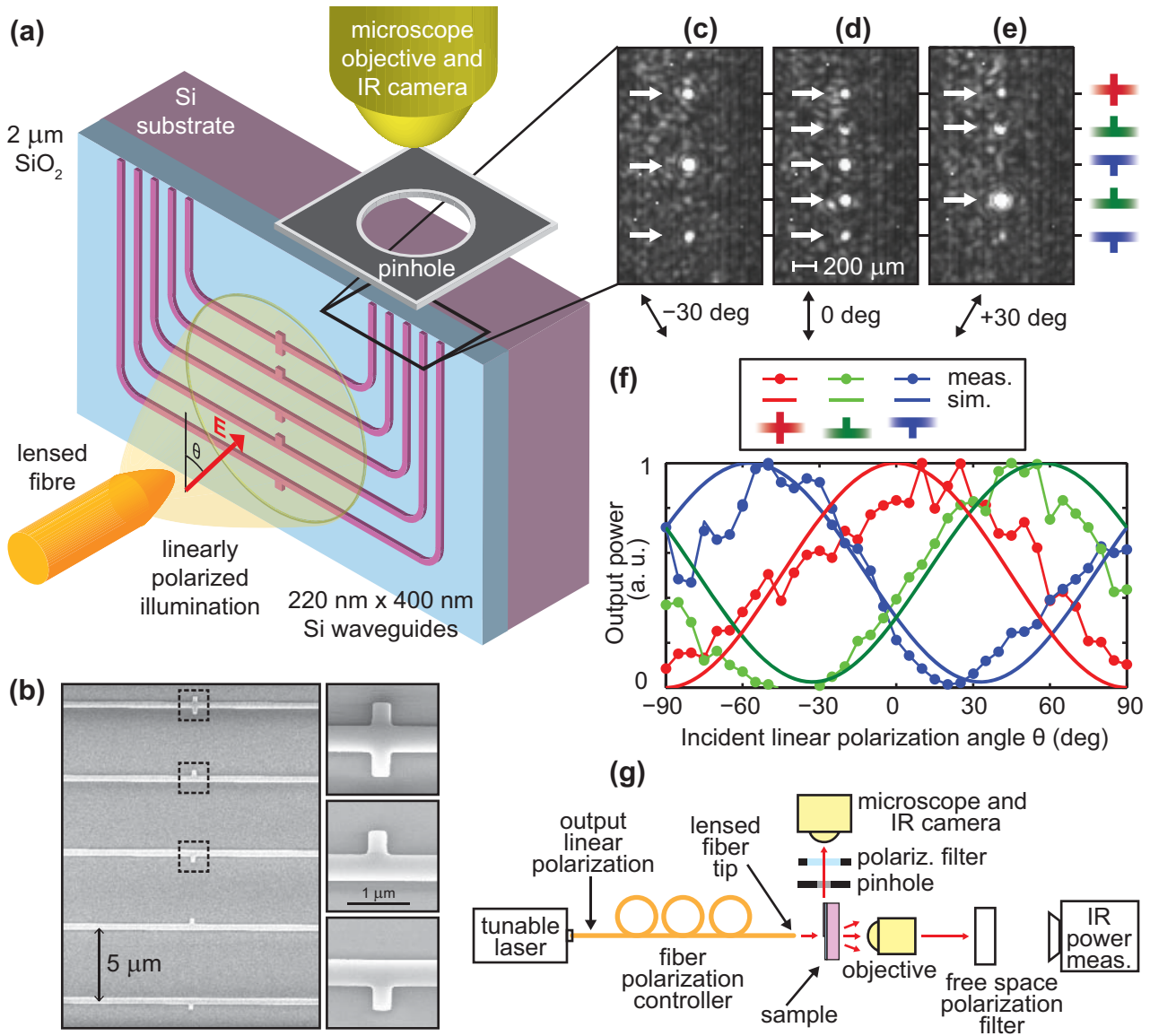

**Fig. 3. Experimental setup and results:** (a) Depiction of the experimental setup (not to scale) (b) Scanning Electron Micrograph (SEM) image of the scatterers in the fabricated and measured sample (c-e) Infrared images of the waveguide outputs captured by the camera for three different linear polarization illumination of the lensed fibre (λ=1550 nm). (f) Measured and simulated output power of the waveguides for different angles of incident linear polarization, obtained by processing the infrared images from the camera. (g) Schematic of the full experimental setup showing the polarization control and monitoring of the incident illumination.

Silicon waveguides have been shown to achieve complex radiation [2,20]. In our case, reciprocity implies that our scatterer can also act as a transmitting antenna with two input silicon waveguides, radiating polarized light in the normal direction with one linear polarization or another depending on the input waveguide used, as shown in our measurements of Fig. 4. See details in Supplemental Materials [19].

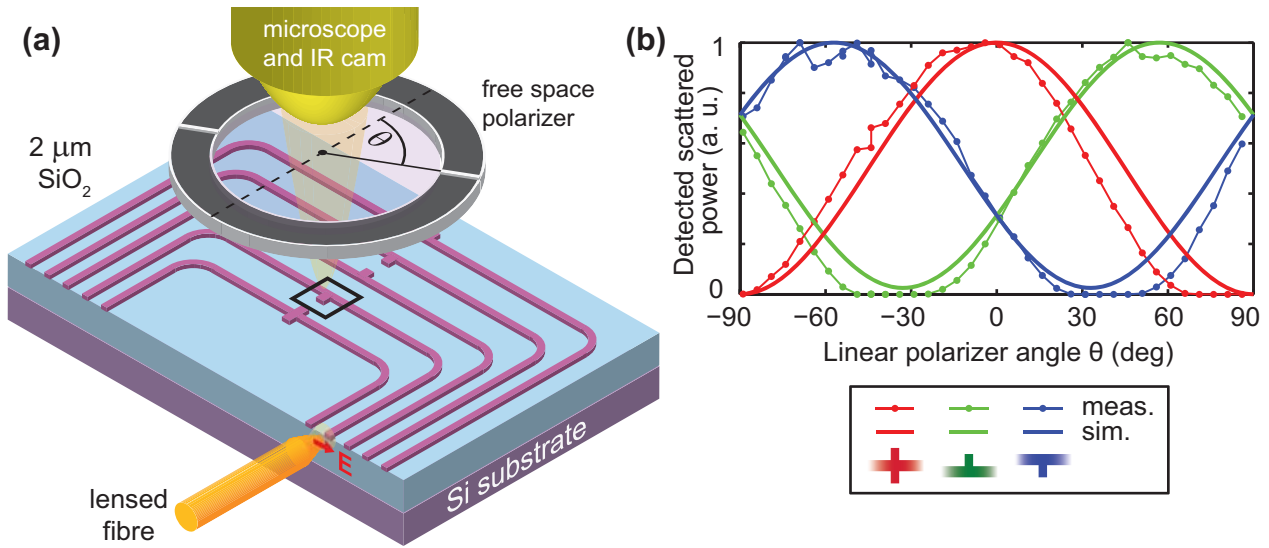

**Fig. 4. Reciprocal experiment: setup and results for an emitting scatterer:** (a) Depiction of the experimental setup (not to scale) (b) Measured and simulated output power of the scattered field component at varying linear polarization angles, obtained by processing the infrared images from the camera ($\lambda$=1550 nm).

Based on local near field interference on a single wavelength-size scatterer, a relatively broad bandwidth and good performance of linear polarization selection is achieved: clearly more complex scatterers with a greater parameter space would allow improved designs and different functionalities, for example, sorting of circularly polarized light could be designed by aiming at $A = \pm iB$. For practical applications, a chain or array of periodically spaced scatterers interfering constructively would increase the effective area of the device, at the cost of a resonant behaviour. Such structure would resemble grating couplers, but in our case, the mechanism results from local field interference instead of being a consequence of diffraction from a periodic arrangement. The scatterer used as a receiver has applications ranging from polarization (de)multiplexing and directional light coupling to polarization-controlled switching and manipulation of polarization-encoded photons in photonic integrated circuits aimed for quantum computing purposes [21,22]. In previous demonstrations of near field interference, free space excitation is used to achieve local field control, but our experiment using the scatterer as a light source turns around this concept, by demonstrating how the near field interaction of a guided mode with a scatterer can achieve a desired polarization in the far field radiation. The measured radiating scatterer could have applications as a dielectric nanoantenna with a controllable polarization in the radiated light.

**Acknowledgements:**

This work has received financial support from Spanish government (contracts Consolider EMET CSD2008-00066 and TEC2011-28664-C02-02). F. J. Rodríguez-Fortuño acknowledges support from grant FPI of GV. D. Puerto acknowledges support from grant Juan de la Cierva (JCI-2010-07479).

**Supplemental Material**

**Materials and Methods**

**Numerical Calculations**– Performed with time domain simulations in CST Microwave Studio$^{TM}$. Open boundary conditions on a 5 μm x 5 μm simulation region with an hexahedral mesh of 30 cells per wavelength where used, and port modes where placed in the two waveguide outputs to measure the complex dispersive A and B for a normally incident plane wave.

**Fabrication**- The structures (see SEM image in Fig. 3(b)) were fabricated on standard silicon-on-insulator (SOI) samples of SOITEC wafers with a top silicon layer thickness of 220 nm (resistivity ρ ~1-10 Ω cm$^{-1}$, with a lightly p-doping of ~$10^{15}$ cm$^{-3}$) and a buried oxide layer thickness of 2 μm. The structure fabrication is based on an electron beam direct writing process performed on a coated 100 nm hydrogen silsesquioxane resist film. This electron beam exposure, performed with a Raith150 tool, was optimized in order to reach the required dimensions employing an acceleration voltage of 30 KeV and an aperture size of 30 μm. After developing the HSQ resist using tetramethylammonium hydroxide as developer, the resist patterns were transferred into the SOI samples employing an also optimized Inductively Coupled Plasma- Reactive Ion Etching process with fluoride gases. Importantly, only a single lithography and etching step is needed to fabricate the structures, which stresses its simplicity.

**Measurements**– To obtain the output power at each waveguide (Fig. 3(f)) we recorded the images on a Xenics Xeva IR camera 1508 mounted on the eyepiece of a 4x microscope objective (National Stereoscopic Microscopes Zoom Models 420 Series), and we processed the images by integrating the camera counts in each output spot. To account for background noise, which imposes a background level of power, we subtract the background power per pixel (obtained from a nearby region without spot) to the spot power per pixel, and we normalize to the background power to compensate for possible variations in the total output power from the fibre (although this step has a negligible effect on the graphs). The polarization at the output of the lensed fibre is varied using a fibre polarization controller. To monitor the output polarization, we make use of a free space polarizer and a Hamamatsu IR camera C2741 used as a power detector behind the silicon substrate (which is transparent to IR) to minimize the polarization orthogonal to the desired one. The complete scheme of the experimental setup is shown in Fig. 3(g). The measurement setup of the scatterer as a transmitter is depicted schematically in Fig. 4(a), the lensed fibre was aligned with each waveguide input to excite the TE mode, and the polarization of the scattered light was analysed with a free space polarizer and the microscope with the IR camera.

**Details of the scatterer used as a transmitter**

The scatterer was initially designed for photon sorting, however, since different linear photon polarizations under normal incidence are sorted into different waveguide directions, reciprocity implies that the radiation of the scattering in the normal direction will have a different linear polarization depending on the input waveguide used. This can be easily understood from an engineering perspective by considering that the scattering matrix of the structure should be reciprocal. The structure can be seen as a system with 4 "ports": the two waveguide ports, and the two normally incident linear polarizations (that can be seen as two different "ports"), resulting in a 4x4 reciprocal scattering-parameter matrix.

When feeding the scatterer with one of the input waveguides, simulations at 1583 nm (corresponding to Fig. 1 in the main text, but used reciprocally) show that the radiation pattern is actually quite complex, showing several lobes in different directions, each with a particular polarization. The radiation in the normal direction is linearly polarized in the predicted angle, and its power is around −14 dB respect to the biggest lobe, which is radiating in a direction at an angle 27º away from the normal and has an angular width at −3 dB of 17º. This lobe, being close to the normal direction, also shows in simulations a radiated light linearly polarized at +40º or −40º with the y-axis, depending on the input waveguide. This means that indeed the scatterer could be used as a relatively effective radiating element with the expected reciprocal properties, even though it wasn't designed for this purpose.

In the measurements, when feeding a single waveguide, our IR camera looking from above detects the scattering of the structure shown in Fig S2. A small scattering is observed all along the waveguide, and an especially high scattering is observed at the curves and output of the waveguide, as expected. The spot coming from the designed scatterer is marked with a circle in Fig. S2. We measured the total power of this spot for the creation of Fig. 4, showing clearly that the emitted light is linearly polarized in the direction expected by reciprocity of Fig. 1.

**Supplemental Figures:**

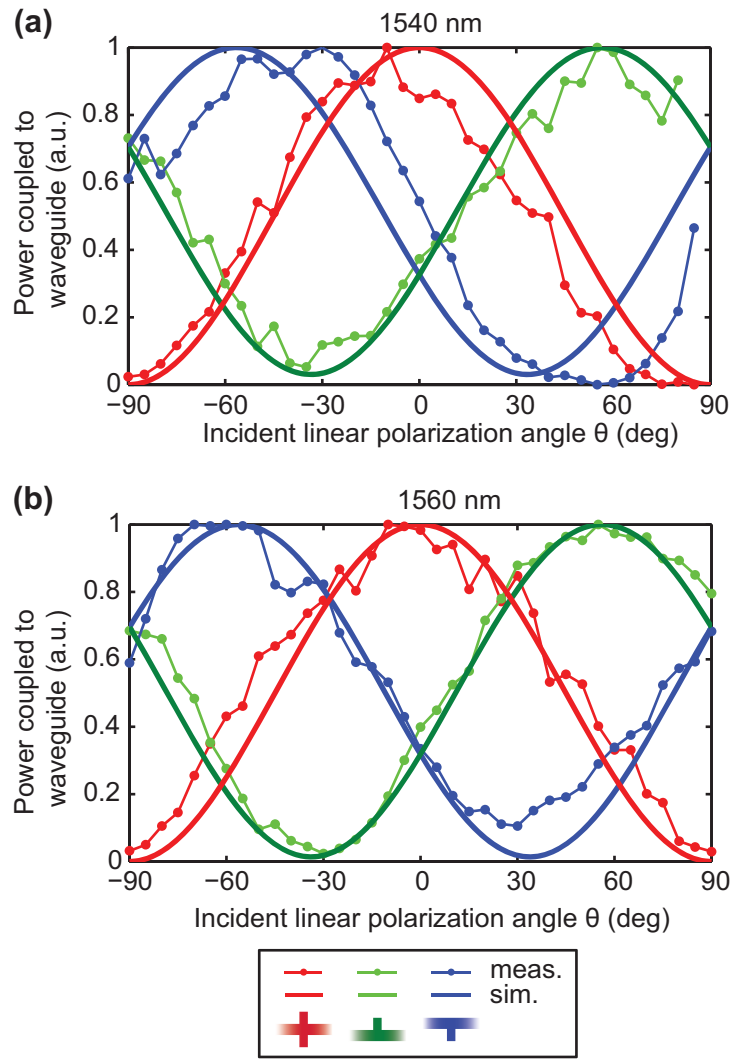

**Fig. S1. Broadband sorting.** Measured and simulated output power of the waveguides for different angles of incident linear polarization at (a) 1540 nm and (b) 1560 nm, obtained by processing the infrared images from the camera at different wavelengths.

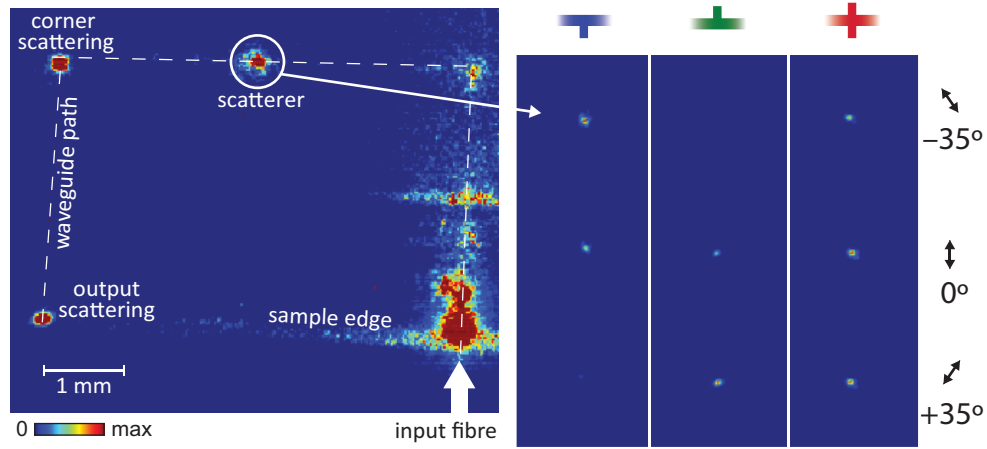

**Fig. S2. Scattering IR image.** (left) Infrared image obtained by an IR camera from the top of a silicon waveguide with a radiating scatterer. The spot measured for the results in Fig. 4 is marked with a circle. (right) IR camera image of the radiating scatterer spot, after a free space linear polarizer at different angles, for three different waveguides being fed. The colour scale is the same for each column. The integration time in the images at the right is reduced to avoid saturation of the image and reduce noise, so the spot of the scatterer is much sharper than the overexposed image at the left.